\documentclass[preprint, superscriptaddress]{revtex4}

\usepackage{graphicx}
\usepackage[utf8]{inputenc}
\usepackage[english]{babel}
\usepackage{amsmath}
\usepackage{subfig}
\usepackage{setspace}

\begin{document}

\title{Insights on unconventional superconductivity in HfV$_2$Ga$_4$ and ScV$_2$Ga$_4$ from first principles electronic structure calculations}

\author{P. P. Ferreira}
\author{F. B. Santos}
\author{A. J. S. Machado}
\affiliation{Escola de Engenharia de Lorena da Universidade de São Paulo, Materials Engineering Department, Lorena -- SP, Brazil}
\author{H. M. Petrilli}
\affiliation{Instituto de Física, Universidade de São Paulo, CP 66318, 05315-970, São Paulo -- SP, Brazil}
\author{L. T. F. Eleno}
\email[Corresponding author: ]{luizeleno@usp.br}
\affiliation{Escola de Engenharia de Lorena da Universidade de São Paulo, Materials Engineering Department, Lorena -- SP, Brazil}

\begin{abstract}

The HfV$_2$Ga$_4$ compound was recently reported to exhibit unusual bulk superconducting properties, with the possibility of multiband behavior. To gain insight into its properties, we performed ab-initio electronic structure calculations based on the Density Functional Theory (DFT).
Our results show that the density of states at the Fermi energy is mainly composed by V--$d$ states. The McMillan formula predicts a superconducting critical temperature ($T_{c}$) of approximately 3.9\,K, in excellent agreement with the experimental value at 4.1\,K, indicating that superconductivity in this new compound may be explained by the electron-phonon mechanism. Calculated valence charge density maps clearly show directional bonding between Hf and V atoms with 1D highly populated V-chains, and some ionic character between Hf--Ga and V--Ga bonds. Finally, we have shown that there are electrons occupying two distinct bands at the Fermi level, with different characters, which supports experimental indications of possible multiband superconductivity. Based on the results, we propose the study of a related compound, ScV$_2$Ga$_4$, showing that it has similar electronic properties, but probably with a higher $T_c$ than HfV$_2$Ga$_4$.

\end{abstract}

\pacs{}

\keywords{HfV2Ga4; ScV2Ga4; DFT;  Multiband superconductivity; Critical temperature}
	
\maketitle

\section{Introduction}

Although superconductivity has attracted the attention of the scientific community for a long time, the understanding of the phenomenon, which  started with the model proposed by Bardeen, Cooper and Schrieffer (BCS) \cite{Bardeen1957}, is still very challenging. The BCS theory, although useful for a large class of superconducting materials, fails to correctly explain other experimentally observed superconducting elements or compounds \cite{Luders2005} and a plethora of different behaviors demands new approaches.

First-principles electronic structure calculations, within the framework of the Density Functional Theory (DFT), has proven to be an important tool to study superconducting materials.
Although strongly correlated systems are beyond the scope of the Kohn-Sham scheme of the DFT, many successful attempts have been made to deal with the description of superconducting 
 materials. In particular, some specific properties of the normal state, e.g. electronic band dispersions and electronic density of states, are very useful to elucidate aspects of the superconducting mechanism and to predict relevant parameters, such as the critical temperature $T_c$ and the elec\-tron-phonon coupling constant $\lambda$. In the last few years, an increasing number of studies appeared using this methodology, either as support for experimental discoveries \cite{sefat2008, chang2016, Machado2017} or fully theoretical investigations \cite{singh2008, subedi2013, tian2016, heil2017}.

Superconductivity was recently {experimentally} reported, by some of the present authors, for the HfV$_2$Ga$_4$ compound, with a critical temperature ($T_c$) of 4.1\,K \cite{Santos2018}. The investigators observed some deviations from the more conventional BCS theory signatures, such as an unusual inflection near $T_c$ in lower and upper critical field as a function of reduced temperature, and a second jump in the specific heat vs. temperature curve. The authors speculated that the experimental results could be either due to sample inhomogeneity or to the presence of more than one superconducting gap at the Fermi surface, resulting in a two-band superconductivity \cite{zehetmayer2013}.

These recent experimental results for the bulk HfV$_2$Ga$_4$ point to a new promising class of materials to study unconventional superconducting behavior. Motivated by these results, here we perform 
 ab-initio electronic structure calculations for HfV$_2$Ga$_4$. We focus our attention on the analysis of the possible mechanisms behind the superconducting properties.
{The theoretical study was extended to a new (possibly) bulk superconducting compound with the same prototype structure, ScV$_2$Ga$_4$}, as a way to manipulate the electronic structure aiming at enhancing the superconducting transition temperature.

\section{Computational Methods}

The ab-initio electronic structure calculations were performed in the framework of the Kohn-Sham scheme \cite{kohn1965} within Density Functional Theory (DFT) using the Full Potential -- Linearized Augmented Plane Wave plus local orbitals (FP-LAPW+lo) method \cite{Singh2006}, as implemented in the WIEN2k computational code \cite{blaha2001}. The Exchange and Correlation (XC) functional was described by the Generalized Gradient Approximation (GGA) in the Perdew-Burke-Ernzer\-hof (PBE) 
  {formulation} \cite{perdew1996}, taking relativistic corrections and spin-orbit coupling (SOC) effects into account. We used muffin-tin spheres with radius $R_\text{MT}=2.0\,a_0$ (Bohr's radius) for all atoms, with $R_\text{MT}\,K_\text{max}$ = 9.0, where $K_\text{max}$ is related to the basis set size {\cite{blaha2001}}. Self-consistent-field (SCF) calculations were carried out with a $32 \times 32 \times 32$ Monkhorst-Pack \cite{monkhorst1976} shifted k-point mesh discretization in the first Brillouin zone. All lattice parameters and internal degrees of freedom 
  were relaxed in order to guarantee a ground state convergence to about 10$^{-5}$\,Ry in the total energy, 10$^{-4}$\,\emph{e} for {electron} density 
   and 0.5\,mRy/$a_0$ for forces acting on the nuclei. The Birch-Murnaghan equation of state \cite{birch1947} was used to fit the total energy as a function of the unit cell volume (keeping $c/a$ constant) at several $c$ values in order to obtain the ground state lattice constants and bulk modulus.
        
   Finally, six different lattice distortions, with 15 intensities for each one (a total of 90 different structures), were used to provide data for the determination of the elastic properties with the ElaStic code \cite{elastic}, using {\sc Quantum Espresso} \cite{Giannozzi2009}  for DFT calculations of deformed structures. The {\sc Quantum Espresso} calculations were performed using PBE SG15 Optimized Norm-Conserving Vanderbilt (ONCV) pseudopotentials \cite{Schlipf2015}, with a cutoff energy of 220\,Ry and 1728 $k$-points in the first Brillouin zone. Anderson's simplified method \cite{anderson1963} was then employed for the calculation of the Debye temperature.
 
\section{Results and Discussion}

\subsection{HfV$_2$Ga$_4$ electronic structure calculations}

\begin{figure}
	\centering
	\includegraphics[width=.6\columnwidth]{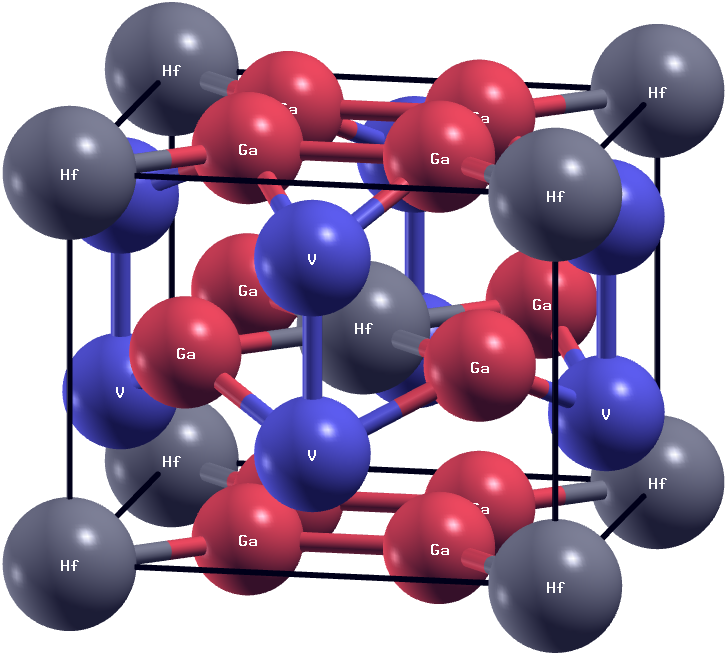}
	\caption{HfV$_2$Ga$_4$ body-centered tetragonal unit cell (conventional setting). Hf (gray), V (blue) and Ga (red) sites are at the 2a (0, 0, 0), 4d (0, 1/2, 1/4) and 8h (0.303, 0.303, 0) Wyckoff positions, respectively.}
	\label{fig:YbMo2Al4}
\end{figure}

HfV$_2$Ga$_4$ crystallize in the YbMo$_2$Al$_4$ prototype (space group \emph{I4/mmm}, Pearson symbol $tI14$), a body-cen\-tered te\-trag\-onal structure composed by a cage-like structure, where Hf sites, at 2a (0, 0, 0) Wyckoff positions, are surrounded by V and Ga sites at 4d (0, 1/2, 1/4) and 8h (0.303, 0.303, 0), respectively \cite{fornasini1976}, as schematically shown in Figure \ref{fig:YbMo2Al4}.

The calculated optimized lattice parameters
are in excellent agreement with the experimental data reported in the literature \cite{Grin1980}, as seen in Table \ref{tab:lat-par}. There is a slight difference of at most 1\% with respect to the experimental values, which is commonly related
 to the inherent imprecision of the approximations required by the computational method \cite{Palumbo2014, Lejaeghere2014, Lejaeghere2016}. The calculated bulk modulus 
  {is} 134.75\,GPa, with a Poisson ratio of 0.24, 
    resulting in 
   416.3\,K for {the} Debye temperature ($\Theta_D$). 
  Our ab initio calculations for $\Theta_D$ reproduce with great accuracy the 
   418.97\,K {value} obtained through experimental measurements \cite{Santos2018}.

\begin{table}
	\caption{Calculated lattice parameters and optimized $8h$ (Ga) atomic position for the HfV$_{2}$Ga$_{4}$ tetragonal compound, compared to experimental values \cite{Grin1980}.}
	\label{tab:lat-par}
	\centering
	\begin{tabular}{lll}
		\hline
		& calc. & exp. \\
		\hline
		$a$, $b$ (\AA) &  6.459&  6.45 \\
		$c$ (\AA) & 5.197 & 5.20 \\
		$8h$ (Ga) & (0.303, 0.303, 0) & (0.303, 0.303, 0) \\
		\hline
	\end{tabular}
\end{table}

\begin{figure*}
	  \centering
		\subfloat[][]{\includegraphics[width=.48\textwidth]{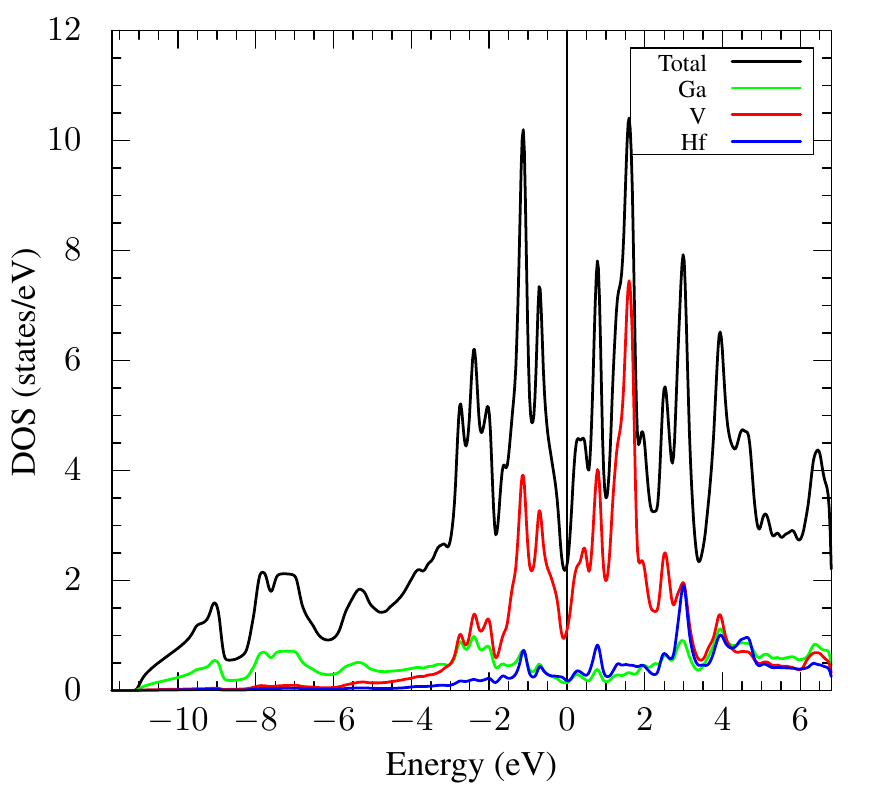}}
		\subfloat[][]{\includegraphics[width=.48\textwidth]{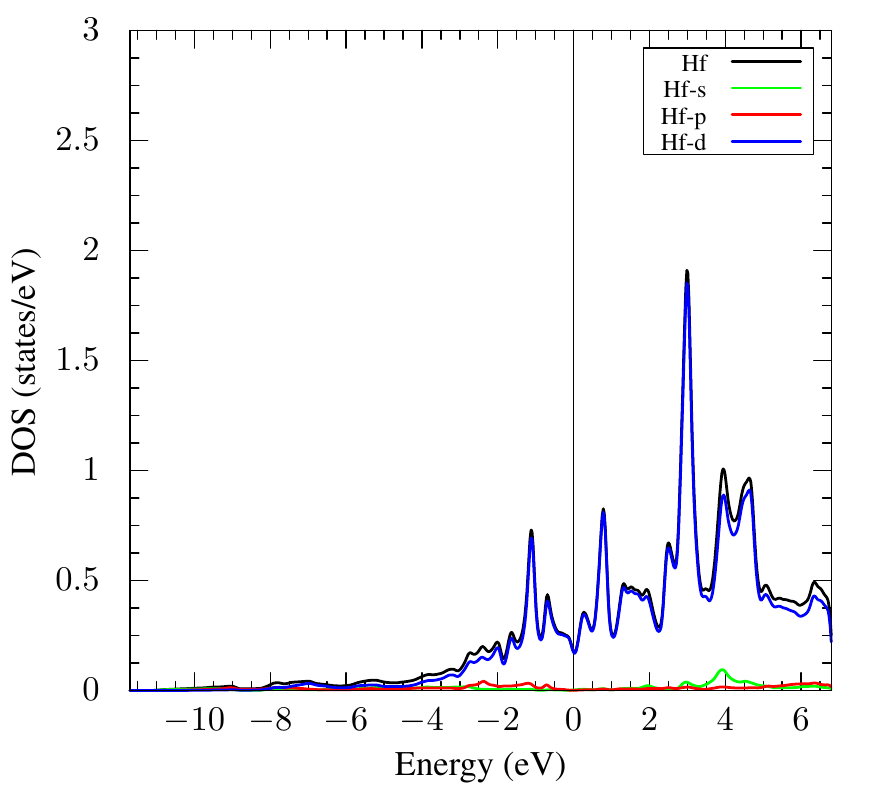}}\\
		\subfloat[][]{\includegraphics[width=.48\textwidth]{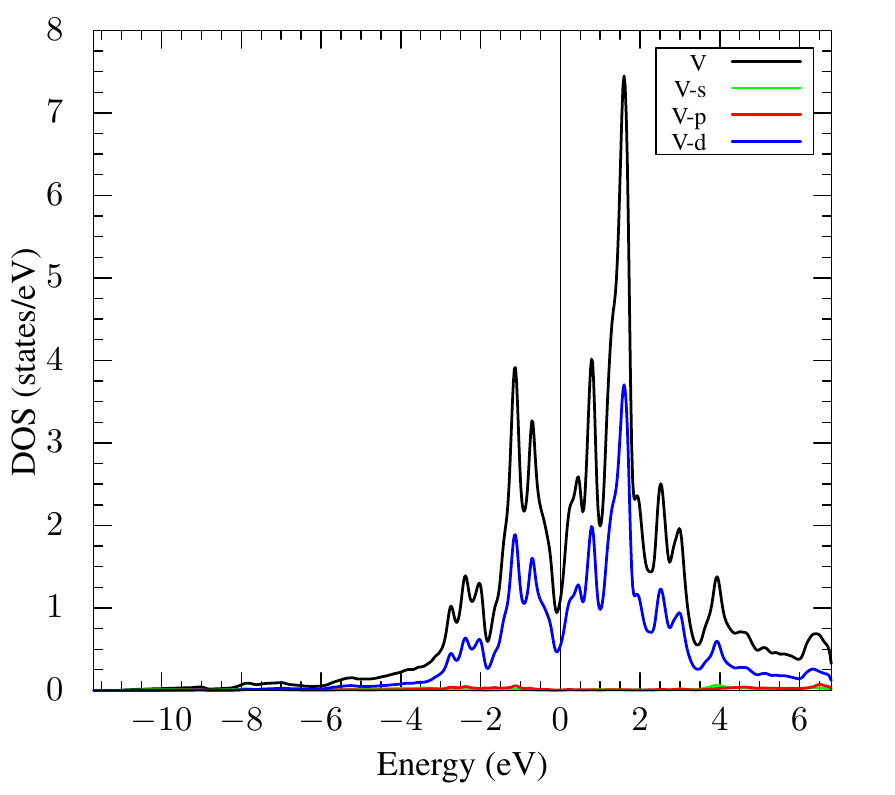}}
		\subfloat[][]{\includegraphics[width=.48\textwidth]{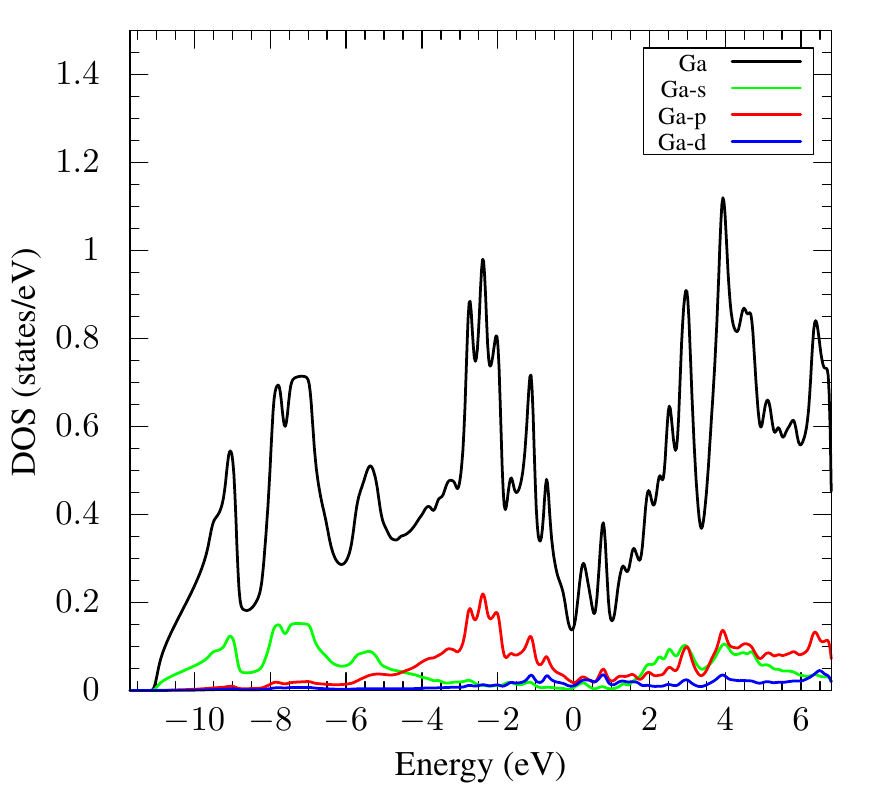}}
		\caption{(color online) (a) Total and site projected density of states for HfV$_2$Ga$_4$. The orbital-projected (\emph{s, p} and \emph{d}) contributions at each site are also shown: (b) Hf, (c) V, and (d) Ga . The Fermi level is set at 0\,eV in all figures.}
		\label{fig:dos_HfV2Ga4}
\end{figure*}

The total density of states (DOS), as well as the site and orbital projected density of states (PDOS), are shown in Figures \ref{fig:dos_HfV2Ga4}a-d. Both occupied and unoccupied states involve considerable hybridization, as seen in Figure \ref{fig:dos_HfV2Ga4}a.
In the lowest energy region  Ga orbitals are dominant, with some contribution from V; in the region around  the Fermi level (from $-2.5$\,eV to 3\,eV), V states are prevailing, mainly due to V-$d$ character (notice the different PDOS scales on Figures \ref{fig:dos_HfV2Ga4}b-d),  with some  Hf-$d$ and Ga-$p$ contributions;  in the higher (above 3\,eV), unoccupied energy region,  Hf and  V states contribute equally.
%
%
Almost half of the total DOS at the Fermi level is due to V, although these states are extended along the whole studied energy region.

The calculated total DOS at the Fermi level is $ N(E_F) = 2.29$\,states/eV. This quantity is related to the linear coefficient of the electronic specific heat $\gamma$, known as Sommerfeld coefficient, given by
\begin{equation}
	\gamma = \dfrac{\pi^2}{3}k_{B}^{2}N(E_F)\,,
	\label{eq:gamma}
\end{equation}
where $k_{B}$ is the Botzmann constant.The calculated $N(E_F)$ leads to a value of 5.41\,mJ\,mol$^{-1}$\,K$^{-2}$ for the theoretical $\gamma_\text{calc}$.
 From the value of the Sommerfeld coefficient $\gamma_\text{calc}$ resulting from the ab-initio calculations and the experimentally measured value ($\gamma_\text{exp}=8.263\,$\,mJ\,mol$^{-1}$\,K$^{-2}$)  \cite{Santos2018},
 we can estimate reasonably well  the elec\-tron-phonon coupling constant $\lambda$ using the well-known approximation \cite{dugdale2011, ram2012}
\begin{equation}
	\lambda = \dfrac{\gamma_\text{exp}}{\gamma_\text{calc}} - 1\,,
	\label{eq:lambda}
\end{equation}
which stems from the fact that the calculations give static (0\,K) results. Following Eq. (\ref{eq:lambda}), we arrive at  $\lambda = 0.53$. This value can be used to calculate the superconducting transition temperature $T_{c}$ using the empirical McMillan formula \cite{mcmillan1968},
\begin{equation}
	T_c = \dfrac{\Theta_D}{1.45}\exp\left[-\frac{1.04(1+\lambda)}{\lambda - \mu^{*}(1 + 0.62\lambda)}\right],
	\label{eq:mcmillan}
\end{equation}
where $\mu^{*}$ is the Coulomb pseudopotential, which measures the strength of the electron-electron Coulomb repulsion \cite{McMillan1965}. A typical value of $\mu^{*}$ is 0.12, as used in many previous works \cite{ram2012, dugdale2011, subedi2008}.
For the HfV$_2$Ga$_4$ compound, using the calculated $\Theta_D$ and the above values for $\lambda$ and $\mu^*$, we arrive at an estimated critical temperature $T_c=3.9\,$K, in excellent agreement with the experimental (4.1\,K) value. This indicates that the electron-phonon interaction may be the mechanism behind superconductivity in HfV$_2$Ga$_4$. The V states dominate the $N(E_F)$ and therefore have the major contribution for the pairing.


\begin{figure}
	\centering
	\subfloat[][]{\includegraphics[width=.6\columnwidth]{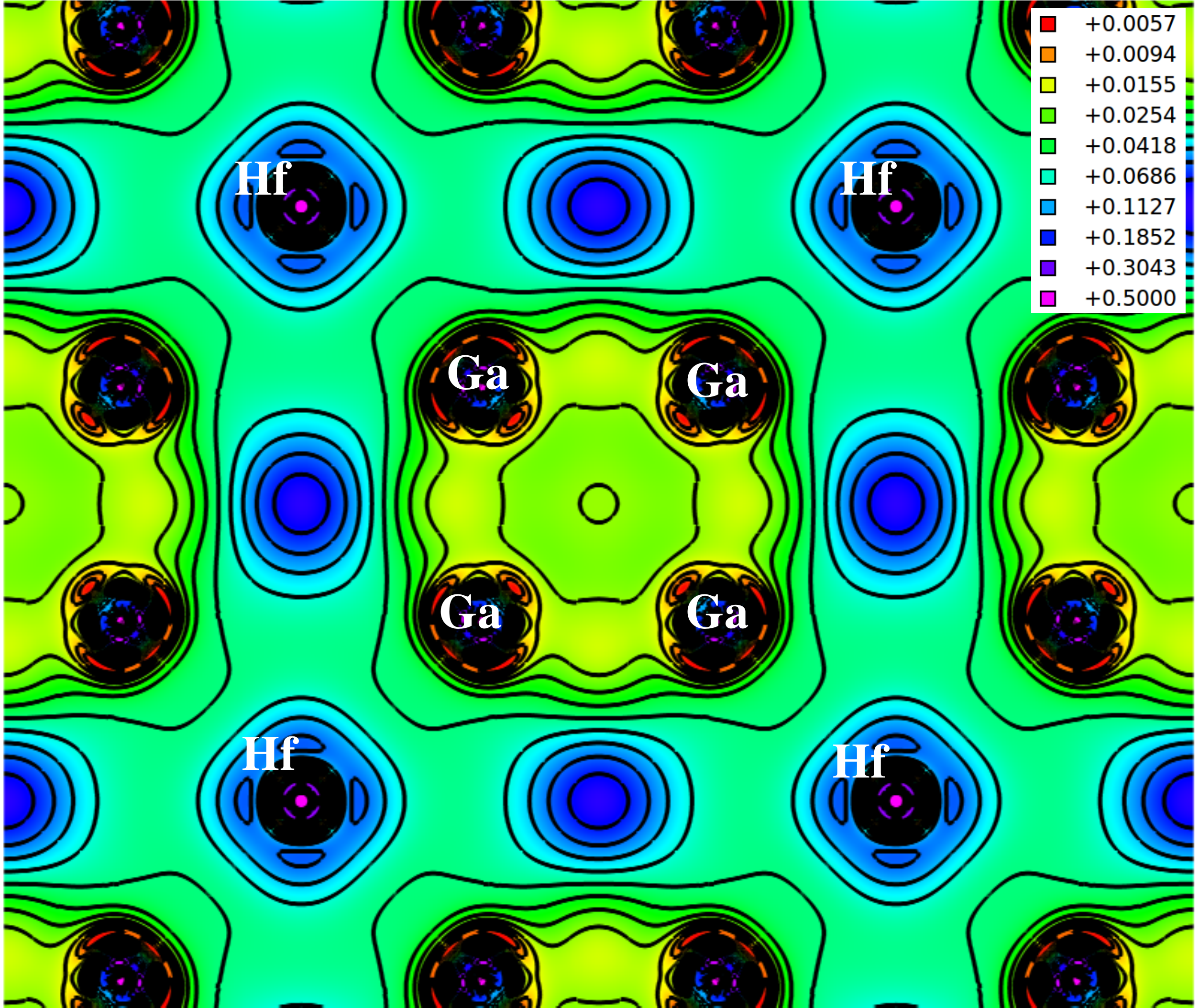}}\\
	\subfloat[][]{\includegraphics[width=.6\columnwidth]{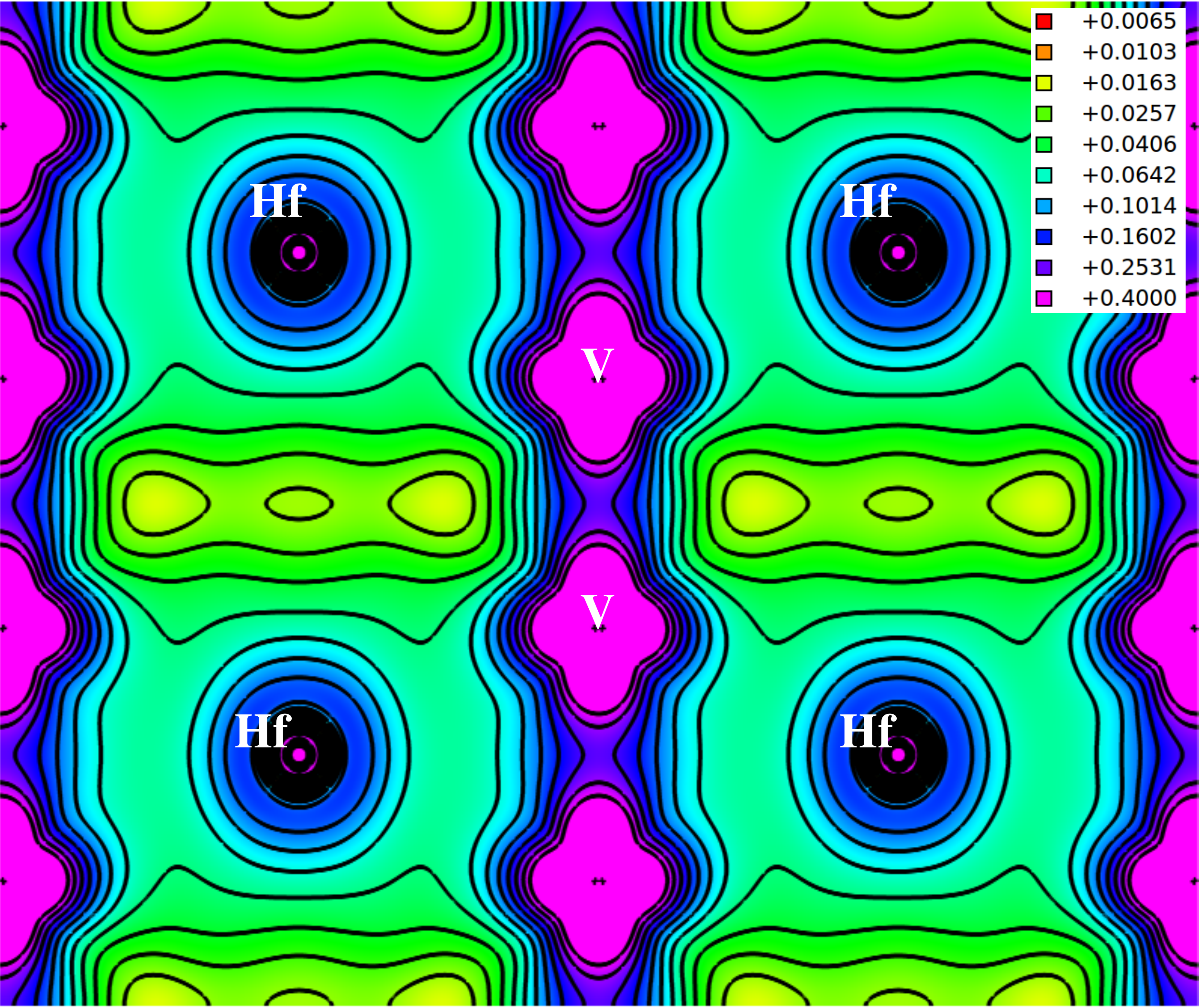}}\\
	\caption{Valence electron density plot in (001) plane (a) and (100) plane (b) for HfV$_2$Ga$_4$. Along a contour the electron density is constant.}
	\label{fig:density_HfV2Ga4}
\end{figure}

The nature of the atomic bonding can be further elucidated with the help of valence electron density plots such as those shown in Figure \ref{fig:density_HfV2Ga4}, in which the electron density is plotted, with an appropriate logarithmic scale in a (001) plane, passing through the center of Hf and Ga nuclei within a unit cell (Figure \ref{fig:density_HfV2Ga4}a), and a (100) plane, passing through the center of Hf and V nuclei (Figure \ref{fig:density_HfV2Ga4}b). 
It should be noted that, in Figure \ref{fig:density_HfV2Ga4}a, the non-labelled high-density regions are V nuclei not centered on the (100) plane.
The plots clearly show a directional shared  bond between Hf and V atoms, evidenced by the density contours in the (100) and (001) planes. This reveal that Hf atoms, which are ``locked''  in the center of a cage-like structure, are not simply passive electron donors: they stabilize the charge transfer to the V atoms (as also observed in Figure \ref{fig:dos_HfV2Ga4}) that, in turn, commands the electronic properties. 

Furthermore, it is interesting to note that the charge density gives rise to a kind of electron sharing channel in the lattice, composed by directional, strongly-bonded, highly populated V chains in the (100) and (010) crystallographic planes.
The Hf nuclei are weakly bonded with the two V atoms within adjacent unit cells in these 1D chains. As a consequence of these V chains that concentrate most of the electronic states that will give rise to Cooper pairs, this electronic configuration may lead to a high anisotropy that could be identified via transport measurements.
Finally, despite the small difference in electronegativity between the atomic species, Hf--Ga and V--Ga bonds exhibit some ionic character. In Figure \ref{fig:density_HfV2Ga4}(a) we can clearly observe isolated clusters containing four Ga atoms within a unit cell, forming weak bonds with adjacent Hf atoms.

\begin{figure*}
	  \centering
		\subfloat[][]{\includegraphics[width=.49\textwidth]{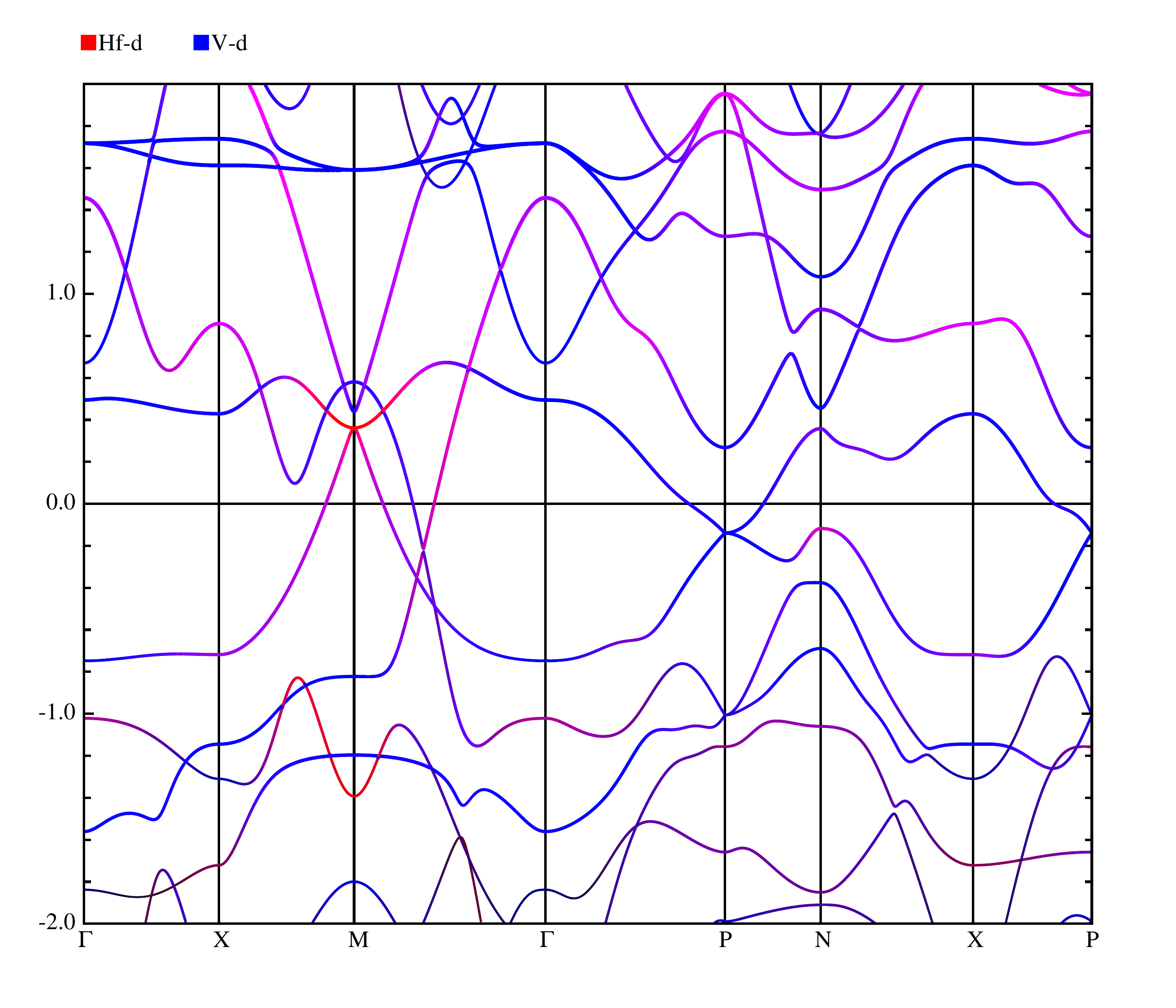}}
		\subfloat[][]{\includegraphics[width=.49\textwidth]{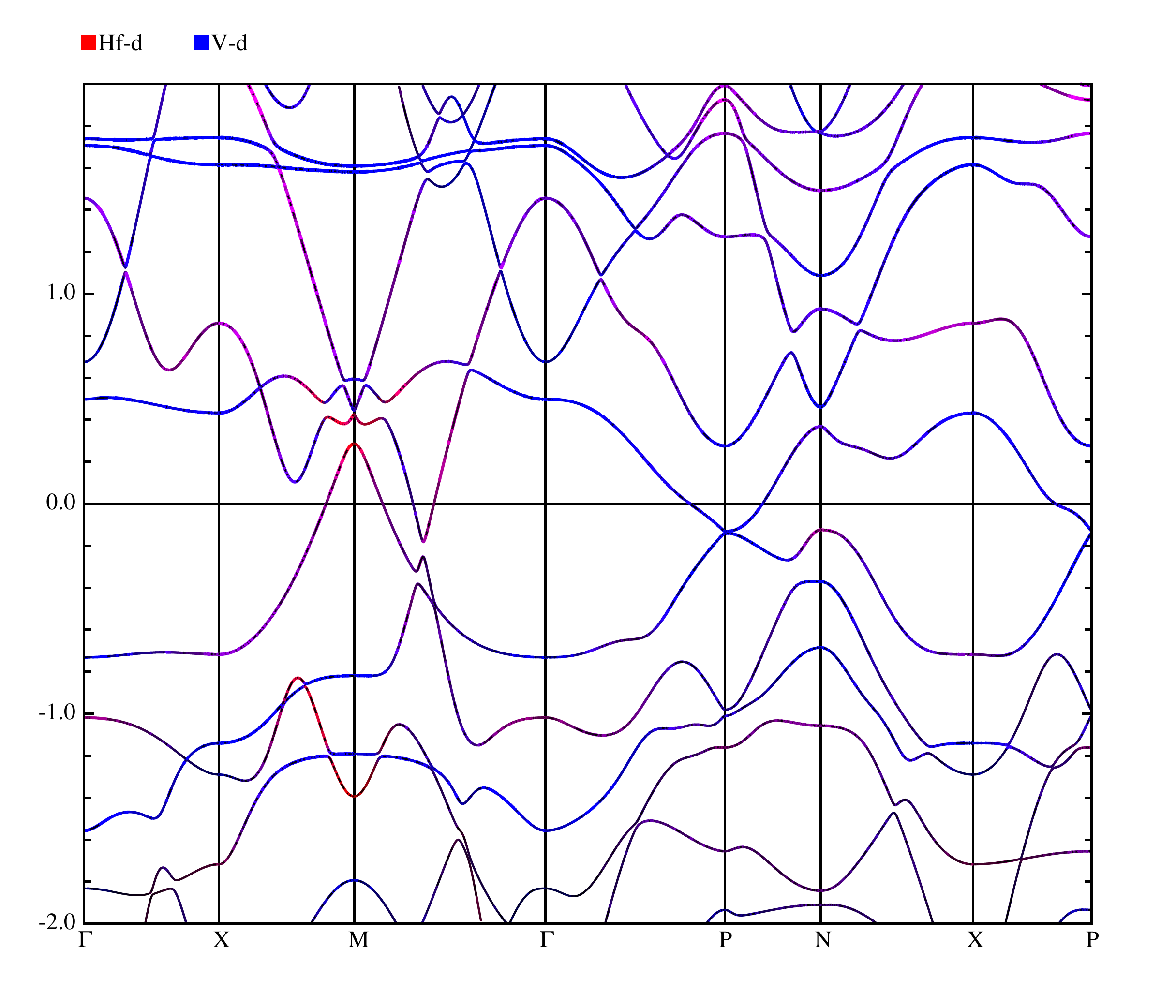}}
		\caption{(color online) Band character plots along high symmetry points in the first Brillouin zone of HfV$_2$Ga$_4$, without (a) and with (b) SOC effects. Colors give a picture about the band character, with color intensity indicating qualitatively the strengh of the contribuition of a given state. Only the Hf-$d$ (red) and  V-$d$ (blue) states  are represented (the Fermi level is set at 0\,eV).}
		\label{fig:band_HfV2Ga4}
\end{figure*}

Figure \ref{fig:band_HfV2Ga4} shows the resulting band character plots along high symmetry points in the first Brillouin zone, not including (Fig. \ref{fig:band_HfV2Ga4}a) and including (Fig. \ref{fig:band_HfV2Ga4}b) spin-orbit coupling (SOC) effects in the calculations. In the band character plots, stronger colors mean stronger character due to the respective orbital projection. Indeed, the cage-like symmetry of the lattice gives rise to complex dispersive metallic bands in the vicinity of the Fermi level. There are two bands crossing the Fermi energy, with very different Hf and V characters. The fact that there are electrons occupying two distinct bands in disconnected sheets of the Fermi surface (corresponding to the two bands crossing the Fermi level)
 supports the experimental evidence of a possible two-gap superconductivity \cite{Santos2018}. These results open a promising scenario for a possible multiband behaviour, so a superconducting gap calculation \cite{Koretsune2017, Giustino2017} would be a very interesting test for that hypothesis.

In the band character plots for HfV$_2$Ga$_4$ shown in Figure \ref{fig:band_HfV2Ga4}, we can see that a hole pocket develops in the M point, 
with a maximum at $\approx\,$0.4\,eV, originated mainly from the Hf-$d$ states containing some mixing with the V-$d$ states. In particular, the electron band crossing the P point
  just below $E_F$ is made up mostly by V-$d$ states. Notice that, near the Fermi level, the band plot unveils dispersive cones with zero gap at M and also along the M--$\Gamma$ direction, as well as one such feature at P. However, when SOC effects are considered, these features are gapped. Indeed, SOC leads to a visible lifting of some band degeneracies, mainly at M and M--$\Gamma$, and less-pronounced at P (just a few meV). Moreover, although these compounds are metallic, SOC broken degeneracy creates a continuous pseudo-gap around the Fermi energy, although the gap almost closes at P (not visible in the scale of Figure \ref{fig:band_HfV2Ga4}b). This kind of signature also occurs in a few nontrivial topological materials like Bi$_{14}$Rh$_3$In$_9$, PbTaSe$_2$ and Cu$_x$ZrTe$_{2-y}$ \cite{rasche2013, ali2014, Machado2017}. Therefore, more detailed experimental and theoretical studies about the possibility of nontrivial topological effects in HfV$_2$Ga$_4$ could be an interesting subject for futures investigations.

\subsection{Theoretical predictions for ScV$_2$Ga$_4$}

Several compounds that crystallize in the same body-centered tetragonal prototype YbMo$_2$Al$_4$, such as RTi$_2$Ga$_4$ (R = Ho, Er, Dy) and RV$_2$Ga$_4$ (R = Sc, Zr, Hf), have been reported in the literature. These compounds are poorly investigated, most efforts having been focused exclusively on magnetic properties in rare-earth compounds \cite{ghosh1993, lofland1994}.

The results reported above for HfV$_2$Ga$_4$ led us to consider an effective way to manipulate the electronic structure of such compounds, aiming at enhancing superconducting properties. In Figure \ref{fig:dos_HfV2Ga4} we can observe that the Fermi level is situated down a deep  valley in the total DOS. As a consequence, the density of states at $E_F$ is extremely sensitive. So, considering a rigid band model, it is a reasonable to assume that an element with a different valence configuration in the $2a$ site of HfV$_2$Ga$_4$ could shift the Fermi level to higher states. Based on what has been presented, we also have carried out first principles electronic structure calculations for ScV$_2$Ga$_4$, to test this hypothesis.

Table \ref{tab:lat-par_ScV2Ga4} shows the relaxed calculated lattice parameters, together with experimental reported values for ScV$_2$Ga$_4$. Following the same methodology applied in the previous section, we reached  $\Theta_D = 447.8\,$K for ScV$_2$Ga$_4$. Unfortunately, in this case, there are no experimental data for comparison.

\begin{table}
	\caption{Calculated lattice parameters and optimized $8h$ (Ga) atomic position for the ScV$_{2}$Ga$_{4}$, compared to experimental values \cite{Grin1980}.}
	\label{tab:lat-par_ScV2Ga4}
	\centering
	\begin{tabular}{lll}
		\hline
		& calc. & exp. \\
		\hline
		$a$, $b$ (\AA) &  6.497&  6.432 \\
		$c$ (\AA) & 5.200 & 5.216 \\
		$8h$ (Ga) & (0.3004, 0.3004, 0) & (0.303, 0.303, 0) \\
		\hline
	\end{tabular}
\end{table}

\begin{figure*}
	  \centering
	    \subfloat[][]{\includegraphics[width=.5\textwidth]{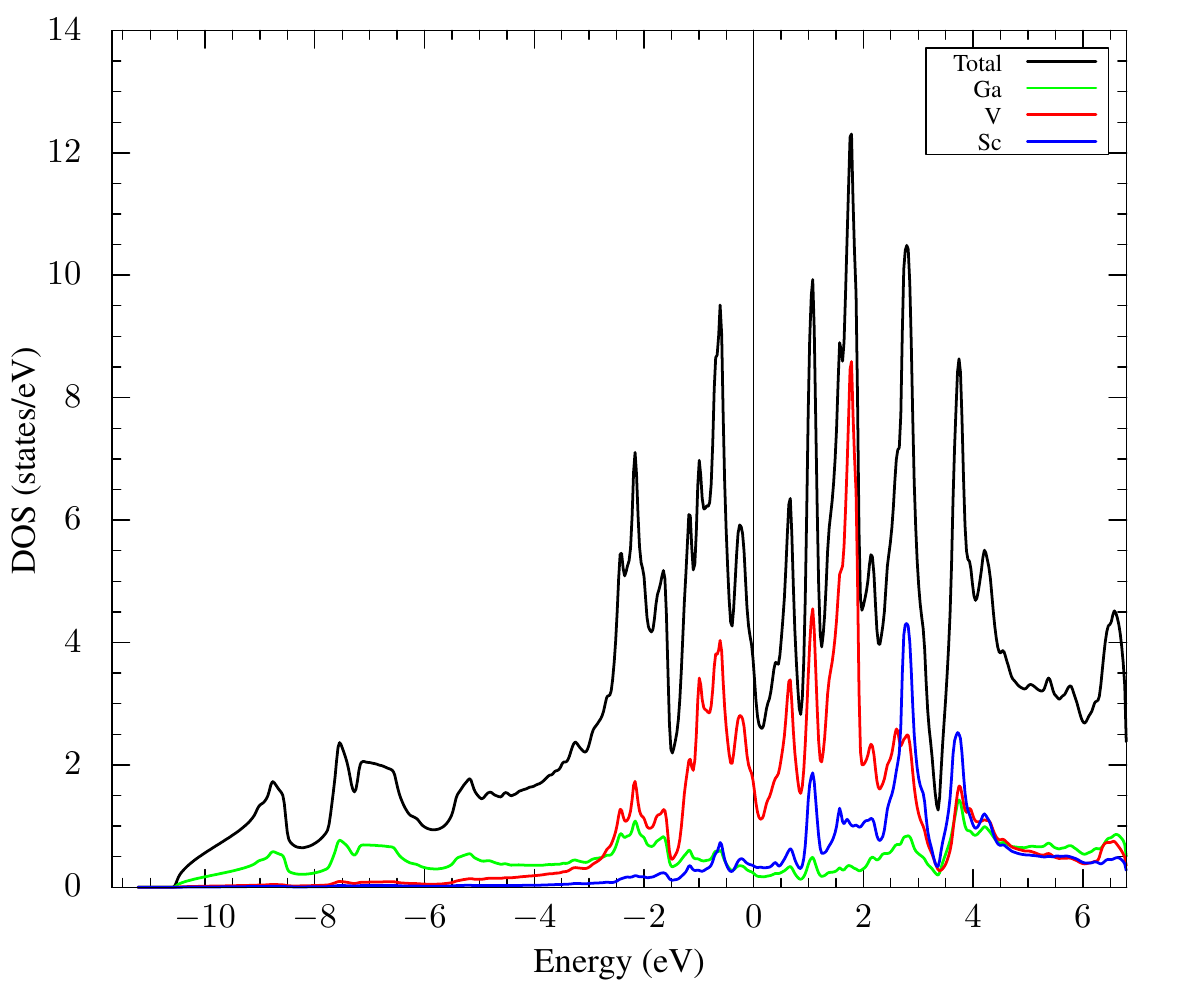}}\\
		\subfloat[][]{\includegraphics[width=.49\textwidth]{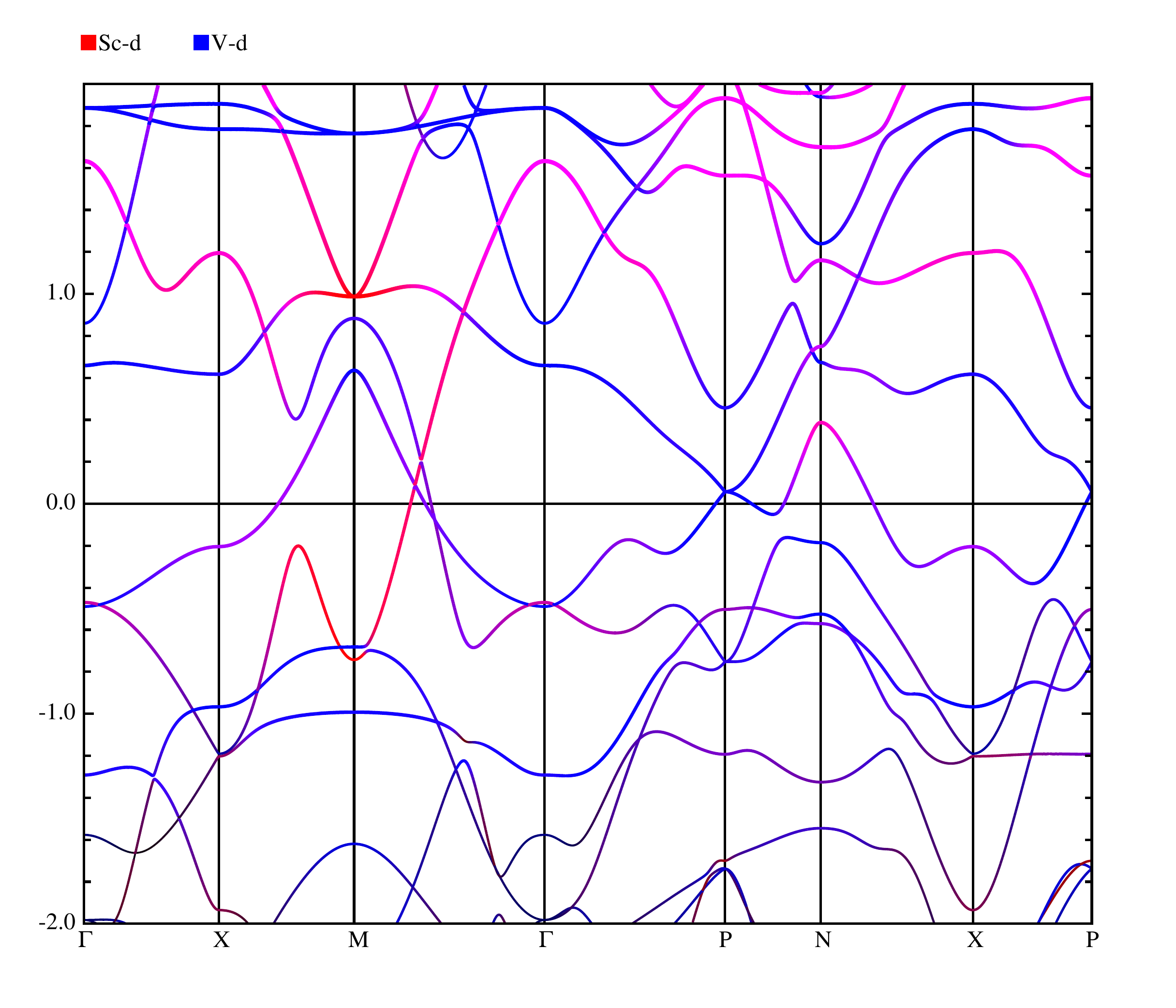}}
		\subfloat[][]{\includegraphics[width=.49\textwidth]{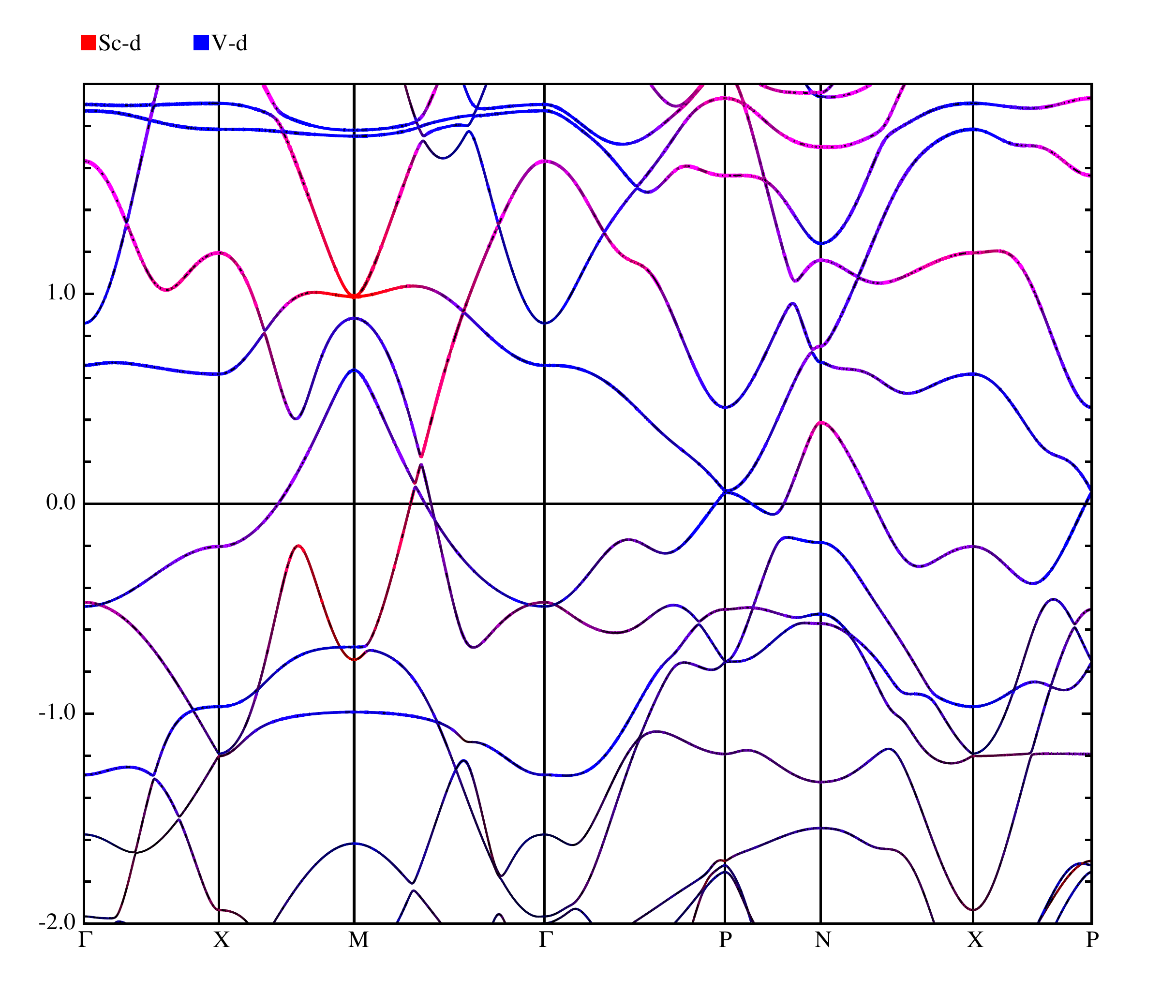}}
		\caption{Total and projected density of states for ScV$_2$Ga$_4$ (a) and band character plots along high symmetry points in first Brillouin zone without (b) and with (c) SOC effects.}
		\label{fig:ScV2Ga4}
\end{figure*}

The nature of atomic bonding is the same showed for the HfV$_2$Ga$_4$ in Figure \ref{fig:density_HfV2Ga4}, with high populated 1D covalent V-chains and Sc atoms acting to stabilize the transfer of charge to the V atoms. Therefore, the DOS overall appearance for HfV$_2$Ga$_4$ is qualitatively identical to ScV$_2$Ga$_4$, as can be verified in Figure \ref{fig:ScV2Ga4}(a). Hence, the contribution of each orbital in the density of states for ScV$_2$Ga$_4$ is also very similar to HfV$_2$Ga$_4$, with a higher contribution due to Sc-$d$ states in the unoccupied bands. The calculated value of $N(E_F)$ is 3.62\,states/eV,  that  
 leads to  $\gamma_\text{calc} = 8.53$\,mJ\,mol$^{-1}$\,K$^{-2}$ using Eq. (\ref{eq:gamma}). Confirming our hypothesis, the presence of Sc atoms instead of Hf in the $2a$ sites causes the Fermi level to shift to a higher DOS value, an increase of about 60\%, escaping from the bottom of the well.

In Figures \ref{fig:ScV2Ga4}(b) and \ref{fig:ScV2Ga4}(c) we show the calculated band structure without and with SOC effects, respectively. It may be seen that the band structure is related to that presented for HfV$_2$Ga$_4$ (Fig. \ref{fig:band_HfV2Ga4}), with similar features in the vicinity of the Fermi energy. Fermi bands in ScV$_2$Ga$_4$ are well described as coming from hybridization between mainly V-$d$ and some Sc-$d$ states. However, SOC in ScV$_2$Ga$_4$ plays only a marginal role, making nontrivial topological effects unlikely. 
 Nevertheless, the important point here resides on the fact that, similar to HfV$_2$Ga$_4$, there are two bands crossing the Fermi level, opening again the possibility for a multiband scenario.

The large contribution of V-d state electrons and the higher DOS value at the Fermi level, attached to the fact that there are electrons originated from two distinct bands in the Fermi surface, strongly suggest that ScV$_2$Ga$_4$ could be a new example of two-band electron-phonon superconducting material with a considerable higher critical temperature than the one reported for the HfV$_2$Ga$_4$ compound.

\section{Conclusions}

In this work we presented ab-initio calculations for the bulk superconductor HfV$_2$Ga$_4$. The McMillan formula predicts a $T_c$ of 3.9\,K, in excellent agreement with experimental reported values (4.1\,K), indicating that superconductivity can be readily explained in an electron-phonon framework. From the signature of the DOS in the vicinity of the Fermi energy, we have proposed to improve the superconducting critical temperature by investigating the ScV$_2$Ga$_4$ compound. Theoretically, we have shown that the presence of Sc instead of Hf in the crystal structure causes the Fermi level to shift to a higher DOS value. The band structure around the Fermi level, which comes mainly from V-$d$ states, and the DOS overall appearance, are qualitatively very similar for both compounds. Valence electron density plots unveil Hf(Sc)-V shared bonding and 1D highly populated V-chains, while Hf(Sc)-Ga and V-Ga bonds have a partially ionic character. It was found that there are electrons derived from two distinct bands in disconnected sheets of the Fermi surface for both compounds, in agreement with the experimental evidence \cite{Santos2018}  of a possible two-gap superconductivity for HfV$_2$Ga$_4$. Finally, we argue that ScV$_2$Ga$_4$ is presumably a new candidate for two-band electron-phonon superconductivity with a higher $T_c$ than HfV$_2$Ga$_4$, a result that should be confirmed experimentally.

\begin{acknowledgements}
	
We gratefully acknowledge the financial support of the Conselho Nacional de Desenvolvimento Científico e Tec\-no\-ló\-gi\-co (Cnpq), Coordenação de Aperfeiçoamento de Pessoal de Nível Superior (Capes), and the Fundação de Amparo à Pes\-qui\-sa do Estado de São Paulo (FAPESP), under Procs. 2017/11023-2, 2016/11774-5, and 2016/11565-7.

\end{acknowledgements}

\begin{singlespace}

\end{singlespace}
	
\end{document}